\documentclass[10pt,journal,compsoc]{IEEEtran}
\usepackage{amsmath, amssymb, amsthm}
\usepackage{graphicx}
\usepackage[a4paper]{geometry}
\usepackage[switch]{lineno}

\input epsf.sty
\newcommand{\includegraphic}[5][,]{%
	\setbox1=\hbox{\includegraphics[#1]{#2}}
	\leavevmode\rlap{\usebox1}
	\rlap{\hspace*{#4}\raisebox{\dimexpr\ht1-#5\baselineskip}{\normalsize{#3}}}
	\phantom{\usebox1}}

\ifCLASSOPTIONcompsoc
\fi

\begin{document}
	
	\title{Resilience in multiplex networks by addition of cross-repulsive links}
	
	\author{Suman Saha,~\IEEEmembership{Project Scientist-I}

		\IEEEcompsocitemizethanks{\IEEEcompsocthanksitem S. Saha is with National Brain Research Centre, NH-8, Manesar, Gurugram-122052, India\protect\\
			E-mail: ecesuman06@gmail.com
		}
		\thanks{}} 
	

	\markboth{IEEE Transactions on Network Science and Engineering}%
	{Saha\MakeLowercase{\textit{}}: Resilience in multiplex networks by addition of cross-repulsive links}
	
	\IEEEtitleabstractindextext{%
		\begin{abstract}
			A multiplex network of identical dynamical units becomes resilient against parameter perturbation by adding selective linear diffusive cross-coupling links. A parameter drift at any instant in one or multiple network nodes can destroy synchrony, causing failure and even collapse in the network performance. We introduced [Phys. Rev. E95, 062204(2017)] a recovery strategy by selective addition of cross-coupling links to save synchrony in the network from the edge of failure due to parameter mismatch (small or large) in any nodes. This concept is extended to  2-layered multiplex networks when the emergent synchrony becomes resilient against a small or large parameter drifting. In addition, the stability of the synchronous state is enhanced from local stability to global stability of synchrony. By the addition of cross-coupling, the network revives complete synchrony in all the nodes except the perturbed nodes, which emerges into a type of generalized synchrony with all the unperturbed nodes. The generalized synchrony is manifested simply by a linear amplitude response in the state variable(s) of the perturbed node(s) by a scaling factor proportional to the mismatch. A set of systematic rules has been derived from the linear flow matrix of the dynamical system representing the nodes' dynamics that helps find the connectivity matrix of the cross-coupling links. Lyapunov function stability condition is used to determine the cross-coupling link strength that, in turn, establishes global stability of synchrony of the multiplex network. We verify the efficacy of our proposed coupling scheme with analytical results and numerical simulations of two examples of multiplex networks. In the first example, we use nonlocal connectivity in each layer with nodal dynamics of the FitzHugh-Nagumo neuron model. Furthermore, we use a second example of a two-layered multiplex network of chaotic R\"ossler systems with random connectivity in both layers. We confirm that our results are generic, independent of nodes' dynamics and network topology.

		\end{abstract}
		
		}

	\maketitle

	\IEEEdisplaynontitleabstractindextext
	
	\IEEEpeerreviewmaketitle

	\section{Introduction}\label{intro}

	\IEEEPARstart{M}{ultilayer} networks \cite{boccaletti2014structure, kenett2015networks, moreno2019, breiger1986cumulated, jalili2017link} evolve in many real-world interactions such as the transport systems consisting of many layers (road, air, and railway) \cite{cardillo2013modeling,wu2020traffic}, human disease network \cite{halu2019multiplex}, ecological network \cite{pilosof2017multilayer}, networks of genes and tissues \cite{liang2019integrating, zitnik2017predicting}, cortical and hippocampal neuronal network \cite{timme2014multiplex} and human brain network \cite{bassett2011dynamic,crofts2016structure}. Multiplex network is a subclass of multilayer networks, where one-to-one interactions between all the nodes in different layers are maintained \cite{mucha2010community,kivela2014multilayer}. In recent time, collective dynamics in multi-layered  networks has been extensively explored \cite{mucha2010community,boccaletti2014structure,sorrentino2012synchronization,sevilla2016inter,nicosia2017collective,leyva2018relay,bergner2012remote, rakshit2019enhancing} including investigations on chimera states \cite{maksimenko2016excitation, majhi2016chimera} and explosive synchronization \cite{khanra2021explosive, jalan2019explosive}. In such multilayer dynamical networks \cite{babloyantz1986low, lytton2008computer, machowski2011power}, synchrony is one of the most important desired states for all practical purposes since it is essential for the efficient performance of the networks. Stable synchrony in a network (single layer, multilayer) is always hindered by internal or external perturbations when the network fails to maintain its desired performance. The best example is the breakdown of a power grid (a multilayer network) due to frequency drift in local nodes, generation, or transmission centers. For efficient functioning, a spontaneous order should sustain that demands resilience of the network synchrony against a disorder (say, a parameter drift). 
	
	\par A network of identical dynamical units \cite{rosenblum1996phase, rakshit2018synchronization} emerges into complete synchrony (CS) for a coupling larger than a critical value \cite{pecora1998master,pecora1997fundamentals} when all the units oscillate in a common rhythm with identical amplitude and phase for all time. On the contrary, for heterogeneity in system parameters (induced or drifting) of a network, CS is broken. At best, a phase synchrony (PS) \cite{rosenblum1996phase,roy2003experimental} may be realized then. In the PS state, the oscillators are phase coherent with almost no amplitude correlation when amplitude distortion of the individual oscillators cannot be avoided. If the system parameters of the dynamical units vary largely, hardly any synchrony can be realized in the network when the individual units oscillate in complete incoherence; even a quenching of oscillation \cite{mirollo1990amplitude, saxena2012amplitude, hens2014diverse, nandan2014transition} can be evidenced, leading to an extreme situation of a collapse of the network performance. The diversity of parameters as heterogeneity in a system, in absence/presence of noise, may even produce a resonance-like phenomenon with amplification of amplitude in an ensemble of oscillators, bistable or excitable \cite{tessone2006diversity}, and limit cycle or chaotic systems \cite{daido2008aging, padmanaban2015amplified}. Similar situations of large parameter variation or drift may occur in multiple nodes in a multilayer network when synchrony can never be achieved, even for large coupling. Therefore, the stability of synchrony as a resilience \cite{gao2016universal} against parameter drift in local nodes of a network is to be established, i.e., the robustness of synchrony against parameter drift is the most desirable. Recently, encouraging reports are coming that create optimism about the constructive role of heterogeneity on synchrony in networks of non-identical dynamical units \cite{zhang2017asymmetry, sugitani2021synchronizing, medeiros2021asymmetry, banerjee2012enhancing}. However, none of the reports addressed how to restore synchrony against a parameter drift. We address the question here to ensure the resilience of synchrony against parameter drifting in vulnerable multiplex networks.
	\par A common strategy of restoring synchrony in networks, in general, is to introduce long-range interactions or links by randomizing the connectivity between any pair of nodes or by rewiring of links \cite{watts1998collective}. A few good benefits of rewiring have been reported in the literature, such as synchrony of dynamical networks \cite{watts1998collective, leyva2017inter, schultz2016tweaking, pade2015improving}, stopping infection progress, and enhancing recovery in an epidemic network \cite{saha2020infection}. A recent study \cite{dwivedi2017optimization} has also proposed optimization of synchrony in static multiplex networks by rewiring. The resilience of synchrony was also realized against significant parameter perturbation by varying the rewiring frequency \cite{rakshit2017time} in time-varying multiplex networks. However, the rewiring of links changes the topology of a network, which is not always allowed or practically feasible in real-world networks of fixed network connectivities. We search here for an alternative strategy to realize the resilience of a multiplex network against significant parameter perturbation in one/multiple nodes in any layer by keeping the original network connectivity intact. 
	
	\par We address the problem in the following manner: Assume a two-layered multiplex network of identical dynamical nodes performing in a stable CS state with its stability conditions determined by the master stability function (MSF) \cite{pecora1998master}, which ensures local stability of synchrony. The critical question is how to stop a breakdown of CS  due to parameter drift in one or multiple nodes in any layer? We propose a strategy to add directed cross-coupling links selectively to the perturbed node(s) of one layer from the nearest/adjacent node of the unperturbed layer and thereby try to prevent the breakdown of synchrony of the multiplex network. The choice of the additional directed link(s) is not arbitrary but made by following a set of systematic rules \cite{saha2017coupling}. Then we define a Lyapunov function of the error dynamics of the network and derive the conditions for establishing the stability of synchrony. Lyapunov function-based stability (LFS)  conditions help derive the cross-coupling strength, establishing globally stable network synchrony. The multiplex network then becomes resilient against parameter perturbation (small or large) \cite{saha2017coupling, saha2017cross}. 
	The cross-coupling links from the unperturbed layer pass additional information to the perturbed layer when the network performance starts deteriorating due to parameter drifting, thereby preventing a network breakdown. 
	\par This approach may have implications in the design of resilient multiplex networks, in general, with its broad applicability in various physical, biological, and engineering \cite{breiger1986cumulated, jalili2017link, cardillo2013modeling, wu2020traffic, halu2019multiplex, liang2019integrating, pilosof2017multilayer, zitnik2017predicting, timme2014multiplex, bassett2011dynamic, crofts2016structure} systems. 
	Although in a different context, the constructive role of cross-coupling has been demonstrated in organic chemical reactions to process novel compounds for use \cite{suzuki1999recent}, ecological species for stabilization \cite{karnatak2014conjugate}, and in the origin of chimera states in complex Ginzburg-Landau systems \cite{sethia2014chimera, hens2015chimera}, other model systems \cite{mishra2015chimeralike}. Here we explore a general procedure of constructing a resilient multiplex network with an analytical description of the choice of self- and cross-coupling and their strength of coupling. The efficacy of our proposition is elaborated with numerical examples of two different network topologies and two model systems as representative dynamical units, one periodic and another chaotic model. It implies that the proposition is independent of nodes' dynamical behavior, to some extent. 
	
	\par Before presenting the technical and analytical details, we brief our main results. We consider first a multiplex network with nodal dynamics represented by the slow-fast FitzHugh-Nagumo (FHN) neuronal model \cite{fitzhugh1961impulses}. The intra-layer coupling is assumed as nonlocal (say, as an example) and unidirectional chemical synaptic type \cite{bazhenov1998cooperative}. The interlayer coupling is diffusive bidirectional, i.e., both the nodes send/receive information to each other simultaneously. The local stability of CS in both the layers is first established using the conditions derived by the MSF \cite{pecora1998master} and considering all the dynamical units of the network as identical. That is, all the nodes follow the same dynamical behavior. Next, a parameter of a single node in one layer is detuned (increased or decreased), which destabilizes CS and destroys the common rhythm. Then a selective, directed diffusive cross-coupling link is added to the perturbed node from the adjacent neighbor of the unperturbed layer. The LFS conditions determine the strength of the cross-coupling link. The addition of the cross-coupling link restores synchrony as emergent generalized synchrony (GS) of the perturbed node with all other nodes while all the other unperturbed nodes remain in the CS state. We clarify later while elaborating the examples on what we mean by a cross-coupling link. The emergent GS \cite{rulkov1995generalized, saha2017coupling, padmanaban2015amplified, boccaletti2000synchronization} as manifested here in the perturbed node is explained here. The amplitude of a state variable of the perturbed node is scaled up/down (amplified/attenuated) by the ratio between detuned parameter and original parameter of the identical nodes, at the same time, phases remain entirely coherent among all nodes. The frequency of the perturbed node thus remains unchanged, which is an essential condition for many real-world networks such as the power grid that does not allow a significant frequency drift.
	Results are successfully verified for perturbation in multiple nodes in one layer. 
	\par  We briefly discuss the design procedure of selecting the coupling links in the Supplementary material (Sec. 1.1, see for some details) and our previous report \cite{saha2017coupling}. The coupling profile or the complete set of coupling functions (self- and cross-coupling) has been identified from a knowledge of the linear flow matrix (LFM) \cite{saha2017coupling} of a nonlinear dynamical system representing the nodal dynamics. The flow of a nonlinear dynamical system, in general, can be separated into its linear and nonlinear components. We extract information from the linear flow matrix of the isolated system that defines the dynamics of a node, how to choose the appropriate self-coupling of the nodes to realize a stable CS  in a network of identical systems. The LFM also tells us the connectivity of the additional cross-coupling links to establish global stability of synchrony. By adopting this procedure, the desired synchrony of the multiplex network can be maintained even when all the nodes of one layer show a distributed heterogeneity. However, each perturbed node needs an additional directed cross-coupling link from the other layer. The perturbed nodes then emerge into a GS state with the unperturbed nodes, which remain in 
	a stable CS state. This phenomenon of emergent GS in the multiplex network as a whole and its resilience against parameter perturbation has further been exemplified in a random multiplex network of chaotic R\"ossler oscillators. Thus the efficacy of our strategy is not restricted to non-locally coupled networks or a particular dynamical system. The coupling profile only changes with the dynamics of the nodes of a network. We emphasize that the multiplex network maintains globally stable CS after adding the cross-coupling links in an identical parametric condition. The network emerges into a globally stable GS under perturbed conditions.

	\section{Multiplex network of FHN systems}\label{sec1}
\begin{figure}[!ht]
		\centering
		\includegraphic[width=1.5in,height=1.5in]{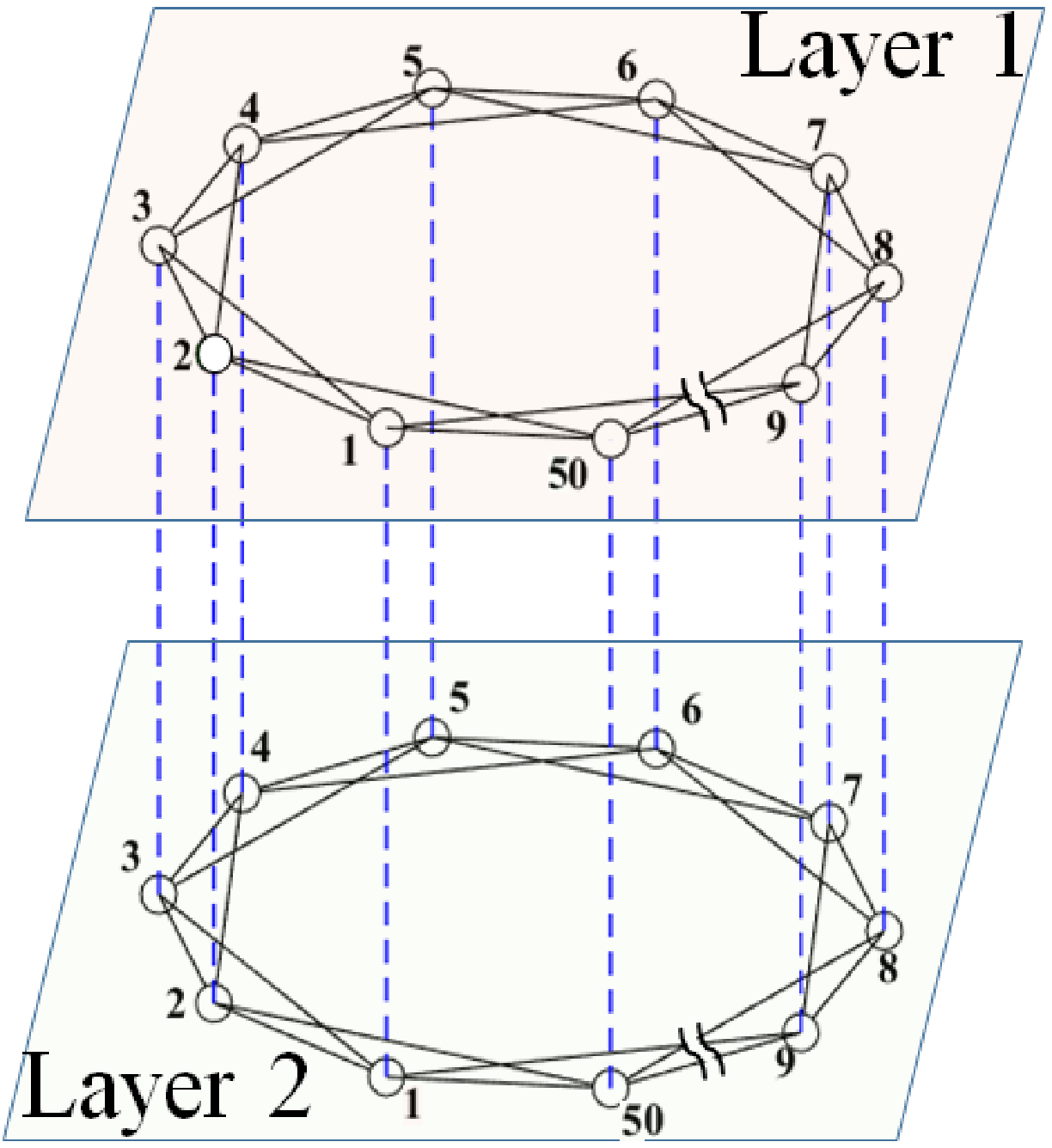}{(a)}{0pt}{0}
		\includegraphic[width=1.75in,height=1.5in]{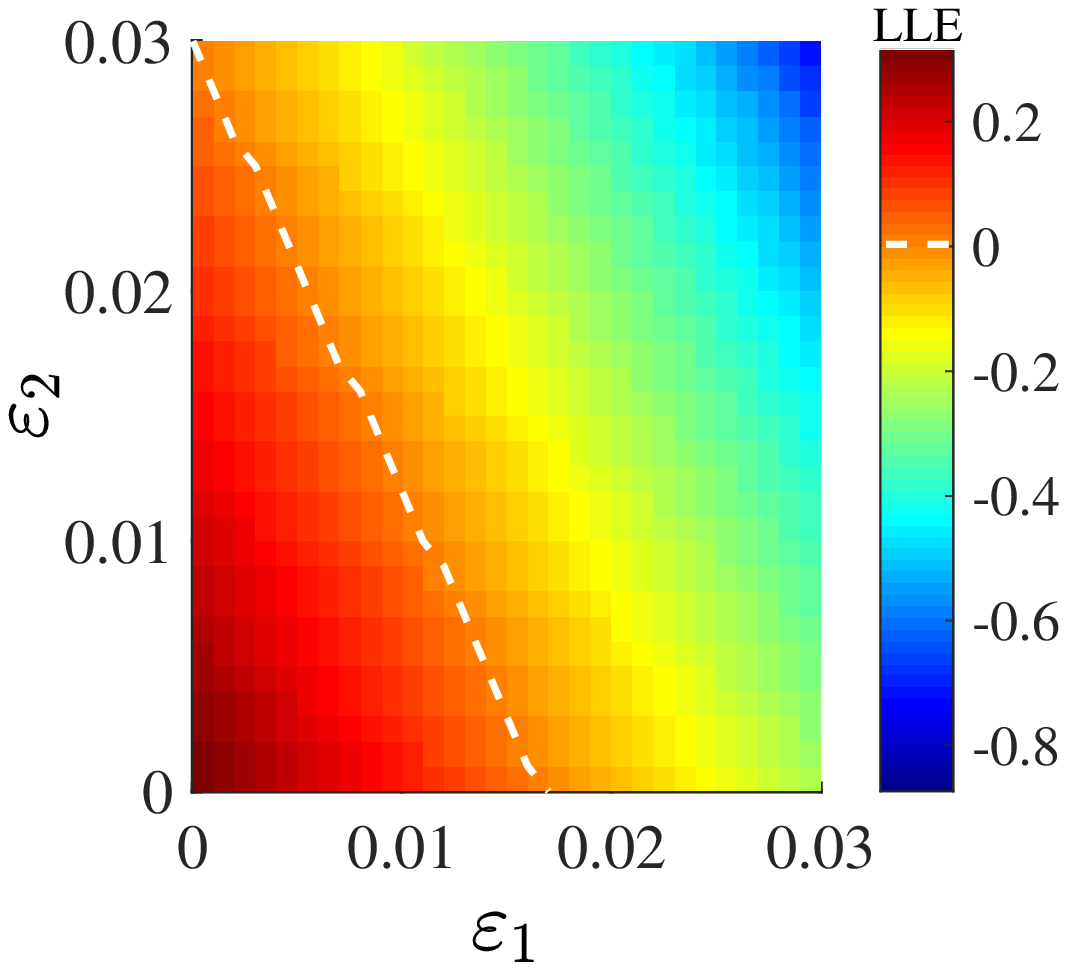}{(b)}{0pt}{0}
		\caption{(a) Schematic diagram of a two-layered multiplex network  of FHN systems. N=50 for both layers (Layer 1 and Layer 2). Each node in both layers is connected to five neighbors ($P=5$) on both sides (black lines) via chemical synaptic self-coupling function. Layer 1 interacts by one-to-one connection to the immediate neighbouring node (dashed blue line) in Layer 2 via diffusive type self-coupling.   (b) Phase diagram of $LLE$ in a $\varepsilon_1-\varepsilon_2$ plane for identical oscillators when cross-coupling links are absent ($\kappa=0$). Color bar shows the finite values of $LLE$, where a white dashed line marks the $LLE=0$ line. The region on the right hand side of the $LLE=0$ line belongs to  locally stable CS state where $LLE<0$. The system parameters are selected as $I=0.4$, $b=0.1$, $V_s=2$.  For identical  case $r^{(1,2)}_i=r=1$, $i=1,\cdots,N$.}
		\label{lle}
	\end{figure}

	Figure~\ref{lle}a shows a multiplex network of two layers where each layer is a non-locally coupled ring of oscillators. One-to-one inter-layer bidirectional interactions are established between the two layers. The solid (black) and dashed (blue) lines indicate the intra- and inter-layer links, respectively. Intra-layer couplings are unidirectional chemical synaptic type (solid black lines) and used for nonlocal  self-coupling \cite{kuramoto2002coexistence, omelchenko2013nonlocal, sakaguchi2006instability} between each node.  The coupling radius is $P=5$ (total neighbors of a node are $2P$) in a layer of $N=50$ nodes (total number of nodes in the network, $2N=100$). The coupling radius $P$ and the network size $N$ may be increased without any change in the results. For this network topology, the dynamics of each node is represented by the slow-fast FHN \cite{fitzhugh1961impulses} system. The coupled systems of  $i^{th}$ node in layer-$l$ is,
	
	\begin{align} \label{eqI50}
		\dot{x}_i^{l} = & F_x(x^{l}_i,y^{l}_i) + \frac{\varepsilon_1 (x^{l}_i-V_s)}{2P}\sum\limits_{j=i-P}^{i+P}  H(x^{l}_j) \nonumber\\
		&+\varepsilon_2 Q(x_i^{(n,l)}-x_i^{(l,n)})+\kappa ~ C_{i,i}^{(n,l)}(y_i^{n}-y_i^{l}), \nonumber \\
		\dot{y}_{i}^{l} = &\;  F_y(x^{l}_i,y^{l}_i,r^{l}_{i}), 
	\end{align}
	
	where $x_i$ and $y_i$ are the state variables of the $i^{th}$ node, $i=1,2,...,N$; $N$ is the number of nodes in both the layers with indices $n,l=1,2$ ($n\neq l$). Subscript indicates the number of nodes in a layer and superscript denotes the number of layers. $V_s$ is a constant bias.
	The flow of intrinsic dynamics of the  $i^{th}$ node in $l^{th}$-layer is described by,
	$F_x(x^{l}_i, y^{l}_i)= x_i^{l} -x^{3l}_{i}-y^{l}_i +I,$ and  $F_y(x^{l}_i,y^{l}_i,r^{l}_{i})=r^{l}_{i}x^{l}_{i}-by^{l}_{i}$. 
	\noindent The intra-layer is nonlocal and the coupling function is represented by $H(x^{l}_j)=\frac{1}{1+e^{-1 0(x^{l}_j+0.25)}}$ as typically used for defining chemical synapses in neurons.  One-to-one inter-layer interactions are bidirectional diffusive type, $Q(x_i^{(n,l)}-x_i^{(l,n)})$. The synaptic coupling
	$H(x^{l}_j)$ and bidirectional $Q(x^{(n,l)}_i)$ are called as  self-coupling functions since they involve the similar variables $x^{n,l}$ and added to the evolution equation of the same variables $x^l_i$. The constants $\varepsilon_1$ and $\varepsilon_2$ represent their respective coupling strengths. The self-coupling establishes locally stable CS in the multiplex network when  $\varepsilon_1$ and $\varepsilon_2$ are larger than their critical values as determined by the MSF, see Figure~\ref{lle}b. On the other hand, the coupling function involving $y^{n,l}$ variables is added to the temporal evolution of the other state variable $x^{l}_i$ with a coupling strength $\kappa$ and henceforth called as cross-coupling links. $C^{(n,l)}_{i,i}$ represents the elements of the connectivity matrix of the directed cross-coupling links from one layer to another layer. The cross-coupling establishes a globally stable synchrony provided the LFS conditions are satisfied. The multiplex network thereby becomes resilient against parameter drifting. The choice of $C^{(n,l)}_{ii}$ is system specific, which follows a set of systematic rules \cite{saha2017coupling} [see Supplementary materials for further details]. The parameters of the FHN system for all the nodes are chosen as $I=0.4$, $b=0.1$, $V_s=2$.  For identical  case $r^{(l)}_i=r=1$. To introduce heterogeneity, the parameter $r^{(l)}_i$ is perturbed from $r=1$ to any other positive value, $r^{(l)}_i>1$;  the perturbation is  randomly introduced within a range $r^{(1)}_i\in [1.5,4]$ as elaborated with an example in Sec.~\ref{sec4}. Results are also true for $r^{(l)}_i<1$, by the same logic, hence not elaborated here.
	
	\par For the realization of CS in an identical multiplex network of the FHN systems, we first derive the critical values of $\varepsilon_1$ and $\varepsilon_2$ (when no cross-coupling is present, $\kappa=0$)  using the MSF (see Supplementary material, Sec. 2). Information on stable CS is captured on a $\varepsilon_1$-$\varepsilon_2$ plane in Figure~\ref{lle}(b) where the onset of CS is marked by a boundary line (white dashed line) when the largest Lyapunov exponent ($LLE$) crosses this line to a negative value. This dashed line defines the critical self-coupling strengths ($\varepsilon_{1c}$ and $\varepsilon_{2c}$): Right-hand side of the white dashed line ($LLE=0$) gives us the range of $\varepsilon_{1}$ and $\varepsilon_{2}$ for the CS regime ($LLE<0$). The color bar indicates the finite values of $LLE$. Note that this CS state based on MSF is locally stable (valid for small perturbation) when the state variables $x_i^l$ and $y_i^l$ in all the nodes in Layer 1 and Layer 2 oscillate with identical amplitude and phase. 
	For a preliminary confirmation of the synchronous state, in the temporal dynamics, we use an error function $e(t)$, considering only $x_i$ variable of each node,
	\begin{align}\label{CE}
		e(t) = & \; \frac{1}{2N} \sqrt{ \sum\limits_{i=1}^{2N}\left[x_i(t)-
			\bar{x}(t)\right]^2},
	\end{align}  
	where, $\bar{x}=\frac{1}{2N}\sum_{i=1}^{2N}x_i$ where $i=1, 2, 3, ...
	2N$ includes all the nodes in two layers. We demonstrate the status of synchrony with time evolution of the state variables of all the nodes and plots of synchronization submanifold in the next section.

	\subsection{Generalized synchrony: Single node parameter mismatch}\label{sec3}
\begin{figure}[!ht]
		\centering
		\includegraphics[width=2.75in,height=1.75in]{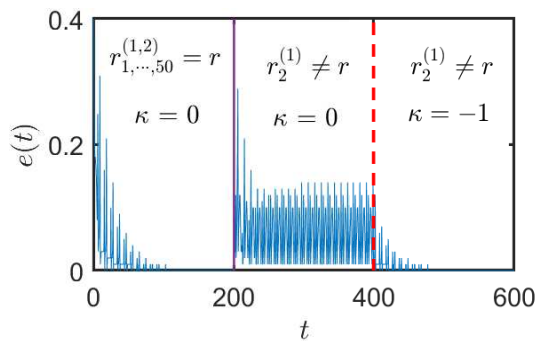}	
		\caption{{\bf Time evolution of the error function.} In identical case ($r^{(1,2)}_i=r=1$), the error function becomes $e(t)$=0 after a transient time,  prior  to any  mismatch ($r^{(1)}_2=2$) introduced in node-2 of layer-1 at $t=200$ (solid vertical line), resulting  in a loss of synchrony as indicated by oscillations in $e(t)$ in the time interval $t=200$ to $400$, in absence of cross-coupling ($\kappa=0$). The synchrony is re-established by the addition of an unidirectional cross-coupling link ($\kappa =-1$) from node-2 in Layer-2 to node-2 of Layer-1 
			at $t \geq400$ (vertical red dashed line). The cross-coupling link restores synchrony in the perturbed network. Self-coupling strengths are $\varepsilon_1=0.1$ and $\varepsilon_2=0.1$, other parameters are $I=0.4$, $b=0.1$, $V_s=2$.}  \label{time_series}
	\end{figure}

\begin{figure}[!ht]
		\centering
		\includegraphics[width=2.6in,height=4.5in]{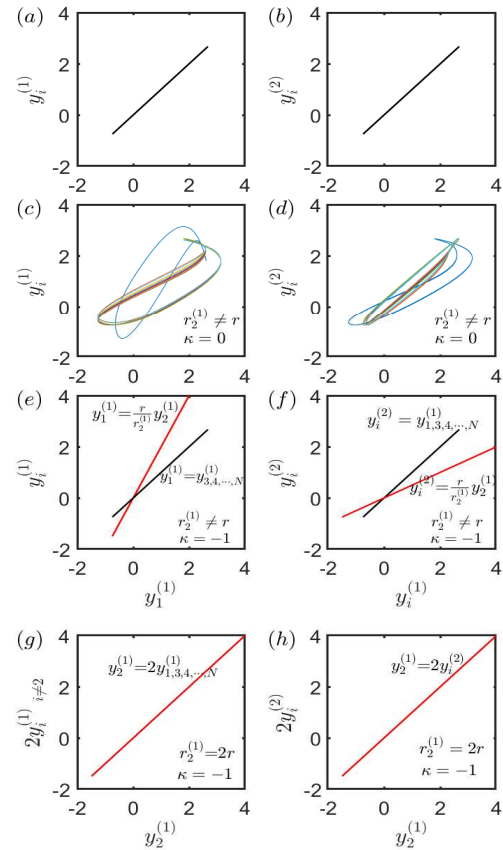}
		\caption{{\bf Intra- and inter-layer synchrony in the multiplex network.} Plot of $y_1^{(1)}$vs. $y_i^{(1)}$ for intra-layer synchrony in (a, c, e) and $y_i^{(1)}$vs. $y_i^{(2)}$ for inter-layer synchrony in (b, d, f). CS state (a, b) is realized  in two layers for $\varepsilon_1$=$\varepsilon_2$=$0.1$, determined by the MSF, in absence of  cross-coupling $\kappa$=$0$. (c, d) Loss of  synchrony is observed due to heterogeneity in node-2 induced by detuning $r^{(1)}_2$=$2$, where  $r^{(1)}_{i}$=$r^{(2)}_{i}$=$r$=$1$ for all other nodes, when $\kappa$=$0$. (e, f) Synchrony is restored by the addition of a cross-coupling link ($\kappa$=$-1$) from node-2 of layer-2 to the perturbed node-2 of layer-1. Addition of cross-coupling link enhances MSF based local stability of synchrony to globally stable synchrony as established by the LFS conditions as detailed in the main text. 
			Node-2 of layer-1 transits to a GS state as indicated by its rotation (red line) from the CS state (black lines) maintained by all other nodes. Amplitude scaling of $y_1^{(1)}$ is indicated by the enlargement of the GS manifold (red line). (g ,h) Actual nature of synchrony of the perturbed node against all other nodes are shown by plotting  $y_2^{(1)}$vs. $2y_i^{(1)}$ and $y_2^{(1)}$vs. $2y_i^{(2)}$,  which also confirms the amplification by a factor of parameter mismatch ratio ($a$=$\frac {r^{(1)}_{2}}{r}=2$). 
		} \label{syn_mani_50}
	\end{figure}

	A stable CS state is realized in the identical multiplex network for a choice of $\varepsilon_1>\varepsilon_{1_c}$=0.011 and $\varepsilon_2>\varepsilon_{2_c}$=0.01 above the critical value  as estimated by the MSF; the broad range of $\varepsilon_1$ and $\varepsilon_2$ is presented in Figure~\ref{lle}b. A plot of the error function $e(t)$ in Figure~\ref{time_series} captures the collective temporal behavior
	of the network before any perturbation of the nodes, which decays to zero after a transient period. At an instant  $t=200$ (marked by a vertical solid line), the parameter $r^{(1)}_2$ of node-2 in Layer-1 is detuned manually keeping rest of the oscillators identical and $\kappa=0$, which destabilizes the synchronous state when  $e(t)\neq 0$ is seen  oscillatory. At the time $t\geq400$ (vertical red dashed line),  one selective, directed cross-coupling link is added from node-2 in Layer-2 to the perturbed node-2 in Layer-1 
	with a coupling strength $\kappa=-1$, resulting in restoration of synchrony of all the nodes of the multiplex network after a transient time when $e(t)=0$. The strength of the additional cross-coupling link ($\kappa=-1$) is analytically obtained by the LFS analysis of two coupled FHN systems (node-2 of Layer-1 and Layer-2), and it is found valid for the multiplex network (see sec.~\ref{appsm} for analytical calculations) and confirmed by numerical results. The procedure to select the particular type of the cross-coupling link is presented in the Supplementary material.
	Now a question arises: What type of synchrony emerges in the multiplex network? The state of synchrony of the multiplex network is scrutinized using plots of the state variables of the perturbed node against the respective state variables of unperturbed nodes in Figure~\ref{syn_mani_50}. The network becomes resilient to parameter drifting in the sense that any change in $r^{(1)}_2$ cannot break the synchrony of the multiplex network.
	
	\par  The status of intra-layer synchrony is shown in  Figures~\ref{syn_mani_50}(a, c, e),  where $y_{1}^{(1)}$ of node-1 ($i=1$) of Layer-1  is plotted against $y_{i}^{(1)}$ of all the nodes in the same layer. The inter-layer synchrony is shown in  Figures~\ref{syn_mani_50}(b, d, f),  where one-to-one correlation is seen between $y_{i}^{(1)}$ of all nodes in Layer-1 against $y_{i}^{(2)}$ of all nodes in Layer-2. For identical oscillators,  CS is thus realized among all the intra- and inter-layer nodes as shown in Figures~\ref{syn_mani_50}(a) and \ref{syn_mani_50}(b), respectively, for $\kappa$=0 and when $\varepsilon_1=0.1$ and $\varepsilon_2=0.1$, which are chosen larger than their critical values as defined by the MSF plot in Fig.~\ref{lle}(b). Both the intra- and inter-layer CS are destroyed by induced heterogeneity in node-2 (by our choice) of Layer-1 as clearly seen in Figures~\ref{syn_mani_50}(c) and \ref{syn_mani_50}(d), respectively. The synchrony is then restored by adding a selective cross-coupling link as shown in Figures~\ref{syn_mani_50}(e) and \ref{syn_mani_50}(f),  where $y_1^{(1)}$ vs. $y_i^{(1)}$ and $y_i^{(1)}$ vs. $y_i^{(2)}$ plots indicate a strong correlation, but rotate (red lines) away from the submanifold of the CS state (black lines) of the identical network. This indicates a new emergent synchronous state in the multiplex network, when the detuned node variable $y_1^{(1)}$ is no more in a CS state with $y_i^{(1,2)}$ of all the  nodes in two layers. 
	The exact nature of synchrony of the perturbed node against all other nodes is revealed by the  $y_2^{(1)}$vs. $2y_i^{(1)}$ and $y_2^{(1)}$vs. $2y_i^{(2)}$ plots (red lines) in Figures~\ref{syn_mani_50}(g) and ~\ref{syn_mani_50}(h), respectively, where the detuned parameter $r_2^{(1)}$=$ar_i^{(1,2)}$=$2r$ ($a=2$, amplification factor) is involved in the rotation of the synchronization submanifold.
	This is manifested by a rotation of the CS manifold $y^{(1,2)}_1=y^{(2)}_2=y^{(1,2)}_3=\cdots=y^{(1,2)}_N$ (black lines) to an emergent synchronization submanifold, $y^{(1)}_2=2y^{(2)}_2=2y^{(1,2)}_1=2y^{(1,2)}_3=\cdots=2y^{(1,2)}_N$ (red lines). While all the state variables in the network remain completely phase coherent, the amplitude of $y_2^{(1)}$ connected to the perturbed node-2 is amplified/attenuated by a scaling factor $a=\frac{r_2^{(1)}}{r_i^{(1,2)}}=2$, which is the ratio of the detuned parameter and the original parameter. We define this emergent state of synchrony as a class of GS state of the whole multiplex network. 
	A most exciting and novel part is that the state variables of all the intra- or inter-layer nodes maintain the CS state except the perturbed node. The synchrony quality is enhanced by adding the particular, directed cross-coupling link establishing resilience against a parameter perturbation. The phenomenon remains independent of the size of the perturbation (small or large). The amount of perturbation cannot destroy GS in the multiplex network when the amplification/attenuation of the particular state variable of the perturbed node only varies linearly following the amount of perturbation with a constant of proportionality $a$. Interestingly, the cross-coupling link does not affect the CS state or sleeps when all the parameters are identical, making the cross-coupling function zero yet making CS globally stable. The cross-coupling starts working or becomes awake when a parameter drift appears in any node, leading to GS of the network, which is globally stable as elaborated in the next section.

	\subsubsection{Analytical Results: Cross-coupling strength in a reduced system} \label{appsm}
	For a given multiplex network under parameter mismatch in one node (simplest case), we only consider the detuned node and its immediate adjacent node from the other layer, which is in CS with all other nodes in both layers of the network. The entire multiplex network is then reduced to two representative nodes, one detuned node, and one unperturbed node, and now we derive the strength ($\kappa$) of the cross-coupling link using the LFS conditions, between the two nodes, node-2 of Layer-1 and node-2 of Layer-2.
	For ease of calculations, we replace the notations of all the variables of the two nodes by
	$x_2^{(1)}$=$x_1$ and $x_2^{(2)}$=$x_2$, and  $y_2^{(1)}$=$y_1$ and $y_2^{(2)}$=$y_2$, and the parameters by $r_2^{(1)}$=$r_1$ and $r_2^{(2)}$=$r_2$. Similarly, discarding the layer indices, we can write the dynamical equations of the two nodes as,
	\begin{align}
		&\dot{x}_{1} = \; x_1 -x_{1}^3-{y}_{1} +I+ \varepsilon(x_{2}-x_{1})+\kappa(y_{2}-y_{1})\nonumber\\ 
		& \dot{y}_{1} = \; r_{1}x_{1}-by_{1}  \nonumber\\
		&\dot{x}_{2} =  \; x_2 -x_{2}^3-{y}_{2} +I+ \varepsilon(x_{1}-x_{2}) \nonumber\\
		& \dot{y}_{2} = \; r_{2}x_{2}-by_{2}  \label{eq2dif}
	\end{align}
	where $(x_{1,2},y_{1,2})$ are state variables,  $\varepsilon$ is self-coupling strength. The conventional diffusive coupling function  $(x_{2,1}-x_{1,2})$ is defined as self-coupling. A parameter mismatch is introduced by taking $r_1$ and $r_2$ parameters as different. A directed cross-coupling link $(y_1-y_2)$ (from one node to the other) is then added to the dynamics of $x_1$ variable with coupling strength $\kappa$. The choice of both the self- and cross-coupling are made following systematic generic rules and detailed in the Supplementary material. \\
	
	\noindent {\bf Case I: For identical systems:}
	
	The LFS is analyzed first to establish globally stable synchrony in two nodes for the identical case, i.e. $r_1=r_2=r$. The error function $\bf e$=$[e_x,e_y]^T=[x_1-x_2,y_1-y_2]^T$ of system Eq.~\eqref{eq2dif} evolves as,
	\begin{align}
		\dot{e}_x = &\;  e_x-\frac{e_x^3}{4}-\frac{3}{4}e_xe_p^2-e_y-2\varepsilon_2e_x-\kappa e_y\nonumber\\
		\dot{e}_y = &\; re_x -be_y \label{s5}
	\end{align}
	where, $e_p=x_1+x_2$ so that $x_1^3-x_2^3=\frac{e_x}{4}(e_x^2+3e_p^2)$.
	For a global stability of  $(e_x=0,e_y=0)$, we define a Lyapunov function, $V(e)=\frac{1}{2}e_x^2+\frac{1}{2}e_y^2$.
	We first check the stability of $x_1=x_2$, separately, by defining a partial Lyapunov function, $V'(e_x)=\frac{1}{2}e_x^2$ when its time derivative is
	\begin{align}
		\dot{V'}(e_x)=&-e_x^2 \left( \frac{3}{4}e_p^2-1+2\varepsilon_2 \right)-\frac{e_x^4}{4} -(1+\kappa)e_ye_x.\label{s6}
	\end{align}
	$\dot V'$($e_x)<0$ provided  $\kappa=-1$ and $\frac{3}{4}e_p^2-1+2\varepsilon_2\geq 0$. 
	To satisfy the condition, we derive the roots of the equation,
	\begin{align}
		\frac{3}{4}e_p^2-1+2\varepsilon_2=0\\
		e_p^2=\frac{4}{3}(1-2\varepsilon_2) \label{s7}
	\end{align}  
	the positive root of $e_p$ is given by
	\begin{align}
		{e_p} =\sqrt{\frac{4}{3}(1-2\varepsilon_2)}. \label{s8}
	\end{align}
	${e_p}$ will now be positive real if $\frac{4}{3}(1-2\varepsilon_2) \geq0$ that implies ${\varepsilon_2}\leq1/2$. Thus 
	$\dot{V'}(e_x)<0$ is satisfied when 
	\begin{align}\label{cond}
		\kappa=-1, \quad \text{and} \quad \varepsilon_2\leq 1/2,
	\end{align}
	
	\noindent this ensures partial stability of $x_1=x_2$.
	Substituting the conditions in \eqref{cond} into  Eq.~\eqref{s5} and assuming identical  systems ($r_1=r_2$), we obtain $\dot{e}_y=-be_y$, when
	\begin{equation} \label{s9}
		\dot{V}(e_x,e_y)=-\frac{e_x^4}{4} - b e_y^2 <0, \quad \varepsilon_2\leq \frac{1}{2} \quad \text{and} \quad \kappa=-1.
	\end{equation}
	
	\noindent Once this LFS condition \eqref{s9} is satisfied, the CS state $x_1=x_2$ and $y_1=y_2$ becomes globally stable in presence of the selective self- and cross-coupling links. \\
	
	\noindent {\bf Case II: For non-identical systems:}
	
	Now we study the effect of heterogeneity on stability of CS by detuning one of the parameters when $r_1\neq r_2$.
	The induced heterogeneity parameters $r_1$ and $r_2$  are not involved in Eq.\eqref{s6}, hence stability  of the  $x_1=x_2$ is still preserved. After detuning $r_{1,2}$, the equation of $\dot{e}_y$ from Eq.~\eqref{s5} becomes,
	\begin{equation}
		\begin{array}{l}
			\dot{e}_y =r_1x_1-r_2x_2-be_y  = (r_1-r_2)x_1-be_y \\ \quad\; 
			=\frac{r_1-r_2}{r_1}(\dot{y}_1+by_1)-be_y, 
		\end{array} \label{s10}
	\end{equation}
	and this leads to
	\begin{align}
		\dot{y}_1&(1-\frac{r_1-r_2}{r_1})-\dot{y}_2
		=-by_1(1-\frac{r_1-r_2}{r_1})+by_2  \nonumber \\
		\dot{y}_1&\frac{r_2}{r_1}-\dot{y}_2=-b(y_1\frac{r_2}{r_1}-y_2).\label{s111}
	\end{align}
	\noindent From Eq.\eqref{s111}, the revised error dynamics is $\dot{e}_y^*=-be_y^*$, where  the modified error function becomes
	\begin{align}
		e_y^*=y_1\frac{r_2}{r_1}-y_2.\label{s11}
	\end{align}
	Rewriting Eqn. \eqref{s11} in actual terms, we have
	\begin{align}
		e_y^*=y_2^{(1)}\frac{r}{r_2^{(1)}}-y_2^{(2)}.\label{s12}
	\end{align}
	\noindent Accordingly, the Lyapunov function is redefined in terms of the modified error functions whose time derivative is
	\begin{equation}
		\dot{V}^{*}(e_x,e_y^*)=-\frac{e_x^4}{4}-be_y^{*2}<0 \label{s13}
	\end{equation}  
	\noindent which ensures a globally stable synchrony, which  basically represents an emergent globally stable GS state as cliamed in the previous section. 
	The manifestation of the GS state is found in the perturbed node,\\
	$$y_2^{(1)}\frac{r}{r_2^{(1)}}=y_2^{(2)}, \;\text{or}\;
	y_2^{(1)}=\frac{r_2^{(1)}}{r} y_2^{(2)}$$
	
	while all other state variables of all the nodes in the multiplex network maintain CS ($x_1=x_2$ and $y_1=y_2$).
	\par Under this stability condition, all other identical nodes emerge into CS and can be treated as if isolated since the self-coupling terms eventually vanish. However, the cross-coupling link in the perturbed node remains active for any instant of time and plays a constructive role for both identical and non-identical cases. Global stability of CS is maintained, in an identical case, beyond the MSF-based local stability, when the cross-coupling function becomes effectively zero or sleeping since $y_2^1$=$y_2^2$. Under the perturbed conditions, the cross-coupling function is awake, and the network emerges into a globally stable GS state while  MSF fails to ensure the stability of synchrony of any form.
	We find that this particular strength of $\kappa$ as analytically derived for two nodes here works perfectly well for an extensive network and for multi-node perturbation to ensure a resilient multiplex network of FHN systems with an emergent globally stable GS. We must mention here that the amplification factor ($a=\frac{r_2}{r_1}$) is a very general effect for any pair of nodes between the two layers for any topology of the multiplex network. The question remains, how this particular choice of cross-coupling link is to be made? The choice of the cross-coupling function is systematically made by a semi-analytic approach as detailed in \cite{saha2017coupling} and elaborated briefly in the Supplementary material. For a particular choice of node dynamics, an appropriate cross-coupling link can always be found to realize globally stable GS in perturbed multiplex networks of various topologies.
	Most importantly, the original topology of the multiplex network remains unchanged. Only one sleeping directed cross-coupling link is added for each perturbed node to the identical network that awakes only when a parameter drift appears. The next section presents numerical results on perturbation in multiple nodes in a layer.

	\subsection{State of coherence: Heterogeneity in  a layer} \label{sec4}
	We numerically demonstrate the efficacy of the cross-coupling strategy in the case of multiple perturbed nodes. All the nodes of one layer (say, Layer-1) are made heterogeneous by a normal distribution of $r^{(1)}_{i}\in (1.5,4]$ with a mean 3 and standard deviation 0.5 where $i=1,2,\cdots, 50$. This range of $r_i$ and its distribution are arbitrarily chosen.
	\begin{figure}[!h]
		\centering
		\includegraphics[width=3.35in,height=2.65in]{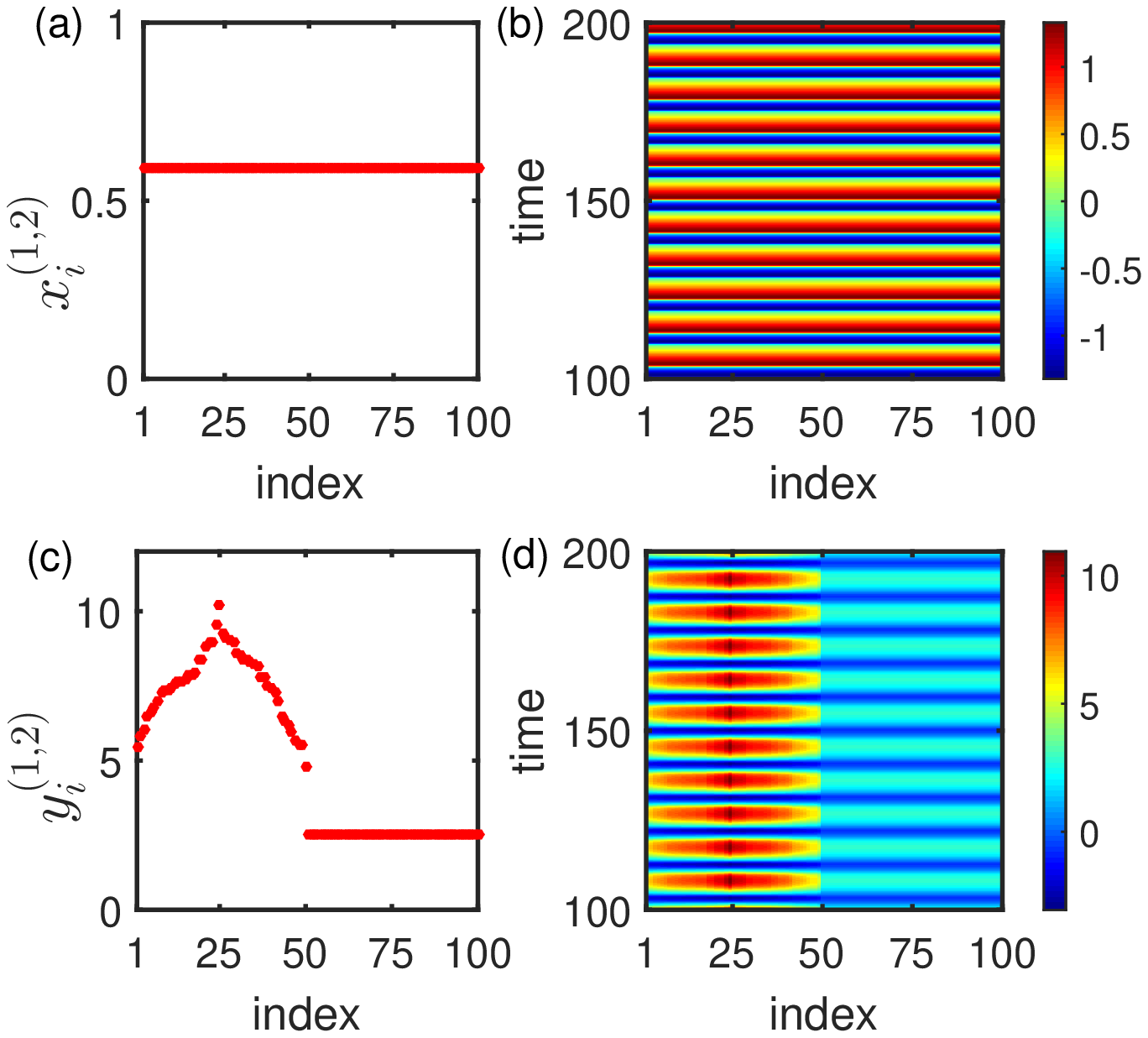}\\
		\includegraphics[width=3.25in,height=1.55in]{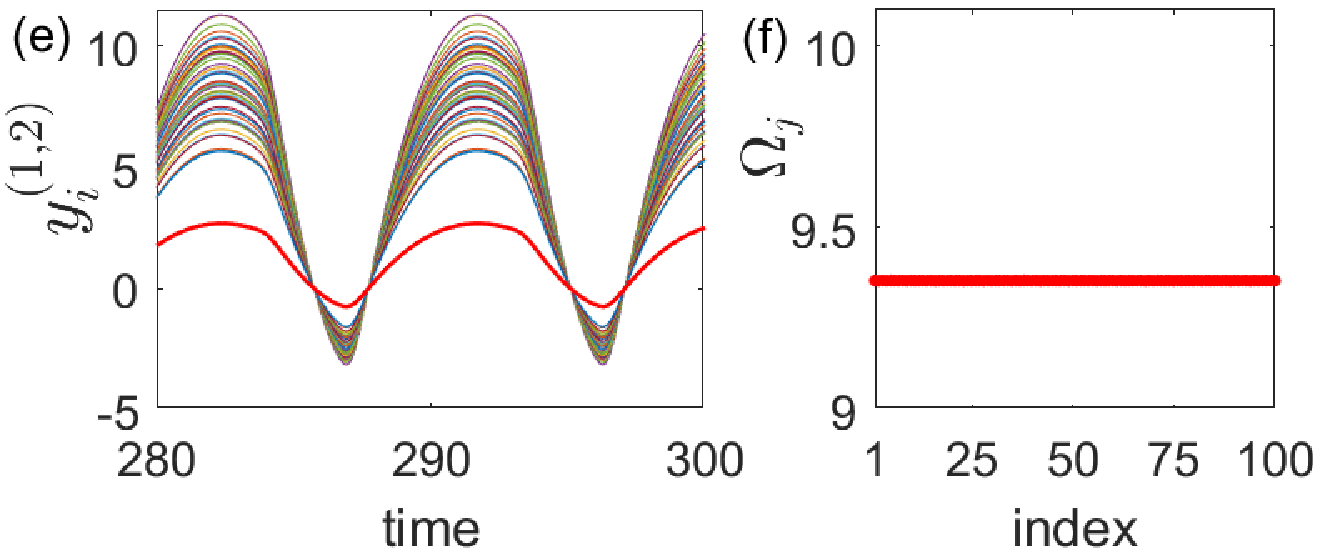}
		\caption{(Color online) {\bf Synchrony in the multiplex network for distributed heterogeneity in one layer.} Parameters $r^{(1)}_{i}$ in layer-1 are detuned randomly within a range of $r^{(1)}_{i} \in (1.5, 4)$ with a  normal distribution of mean 3 and standard deviation 0.5; for unperturbed nodes ($i=51$ to $100$) in Layer-2, $r_i^{(2)}$=$r$=1. (a) Snapshot of temporal and (b) spatio-temporal dynamics of $x_i$ of all the nodes, show CS in amplitude and phase. (c) Snapshot of temporal evolution  and (d) spatio-temporal plots of $y_i^{(1,2)}$ confirms amplified response all perturbed nodes in Layer-1 ($i=1$ to $50$) while all the identical nodes ($i=51$ to $100$) in Layer-2 maintain CS. (e) Time evolution of $y_i^{(1)}$ corresponding to the perturbed nodes in Layer-1 are shown in various colors, which shows phase coherent evolution with varying amplitude. A plot of $y_i^{(2)}$ of all the unperturbed identical nodes in Layer-2 is shown (red line) for comparison and amplitude response of the perturbed nodes. The scaling up of amplitude in $y_i^{(1)}$ is due to distributed heterogeneity in the parameter $r^{(1)}_{i}$. The amplified response in the perturbed nodes is dictated by  the heterogeneity of the nodes (various colors) quantified by the ratio of the parameter mismatch, $r^{(1)}_{i}/r$.  
			(f) Mean phase velocity ($\Omega_{j}$)  confirms complete phase coherence in all the nodes in the network. The frequency of all the nodes remains identical and is preserved against parameter drifting. \label{dis}}
	\end{figure}
	
	The parameter values $r^{(2)}_i$ of all the nodes in Layer-2 are assumed identical i.e., $r^{(2)}_i=r=1$, other system parameters are kept identical too. Unidirectional cross-coupling links from Layer-2 to all adjacent perturbed nodes in Layer-1 are added, keeping other coupling functions and the original topology of the network unchanged. The selection process of the coupling profile (self- and cross-coupling) remains the same as done for single node perturbation. The self-coupling strengths are chosen once again as $\varepsilon_1=0.1$ and $\varepsilon_2=0.1$ larger than the critical values defined by the MSF. The strength of the cross-coupling links is chosen $\kappa=-1$, as derived by the LFS condition for two coupled systems as shown in sec.~\ref{appsm}. We show that our strategy successfully works for perturbation in multiple nodes in the network. 
	\par The emergent behavior of the network under perturbation is described in Figure~\ref{dis}. The CS state in $x_i$ variable of both the layers is not disturbed as shown by a snapshot of $x_i^{(1,2)}$ and their spatio-temporal dynamics in Figs.~\ref{dis}(a) and \ref{dis}(b), respectively. The amplitude of the $y_i$-variables of all the nodes in Layer-1 are amplified by the ratio, $a_i=\left(\frac{r^{(1)}_i}{r}\right)$,  where $a_i$ follows the distribution of $r_i^{(1)}$ of the perturbed nodes and $r=r^{(2)}_i=1$, $i=1,2,\cdots,50$. A snapshot of $y^{(1,2)}_i$-variables in Figure~\ref{dis}(c) shows a pattern of distributed amplitude in the nodes of Layer-1 (1 to 50), while the rest of the nodes (51 to 100) of Layer-2 have identical amplitude. The distribution of amplitude in the first 50 nodes reflects the distribution of $a_i$ due to distribution in $r_i^{(1)}$ in layer-1. The spatio-temporal dynamics in Figure~\ref{dis}(d) corroborates the fact that the first 50 nodes in Layer-1 are phase coherent but with a distribution of amplitude (color bar indicates the distribution in amplitude) while the rest of the 50 nodes in Layer-2 are both amplitude and phase coherent (coherent pattern in cyan color). The amplitude response of the nodes in Layer-1 reveals a distortion-free amplification in the heterogeneous nodes, as confirmed by the temporal dynamics of $y_i^{(1,2)}$ (color lines) in Figure~\ref{dis}(e).  
	To estimate the local phase correlation of all the nodes in Layer-1 and layer-2, a long time average of phase velocity \cite{dana2019chimera} of  $j^{th}$ oscillators is taken,
	\begin{align} \label{phs_velocity}
		\Omega_{j}= \frac{2 \pi M_j}{\Delta t},~~~~j=1,2,\cdots,2N,
	\end{align}
	where $M_j$ is the average number of periods of the $j^{th}$ node for a long time interval $\Delta t$. 
	All the nodes in the two layers are seen completely phase coherent as seen in Figure~\ref{dis}(f), where they all have identical mean frequency $\Omega_j$. It is an essential requirement for many real-world multiplex networks, where any frequency drift is prevented by these additional cross-coupling links against parameter perturbation in any node. The design of cross-coupling links follows a systematic semi-analytical technique as usual, and hence it is quite a novel proposition for realizing a resilient multiplex network.

	\section{Random multiplex network: R\"ossler system} \label{sec5}
	Now we confirm that our coupling strategy is independent of both the connectivity matrices of the layers of the multiplex network and their nodal dynamics. A random multiplex network of two layers is considered with nodal dynamics of a chaotic R\"ossler oscillator and a distributed parameter in one layer.
	The governing dynamics of the $i^{th}$ node,
	{ \small
		\begin{align} \label{eq_ros}
			&\dot{x}_i^{(1)} =  -y^{(1)}_i-z^{(1)}_i + \kappa (z^{(2)}_i-z^{(1)}_i)   \nonumber\\
			&\dot{x}_i^{(2)} =  -y^{(2)}_i-z^{(2)}_i   \nonumber\\
			&\dot{y}_{i}^{(1,2)} = \; x^{(1,2)}_i + a y^{(1,2)}_i 
			+\varepsilon_1 \sum\limits_{j=1}^N A^{(1,2)}_{ij} (y^{(1,2)}_j-y^{(1,2)}_i) \nonumber \\
			& \quad \quad \quad + \varepsilon_2 (y^{(1,2)}_i-y^{(2,1)}_i) \nonumber \\
			&\dot{z}_{i}^{(1,2)} = \; b^{(1,2)}_i -c z^{(1,2)}_i + x^{(1,2)}_iz^{(1,2)}_i,
		\end{align}
	}
	\begin{figure}[!h]
		\centering
		\includegraphics[width=3.5in,height=2.75in]{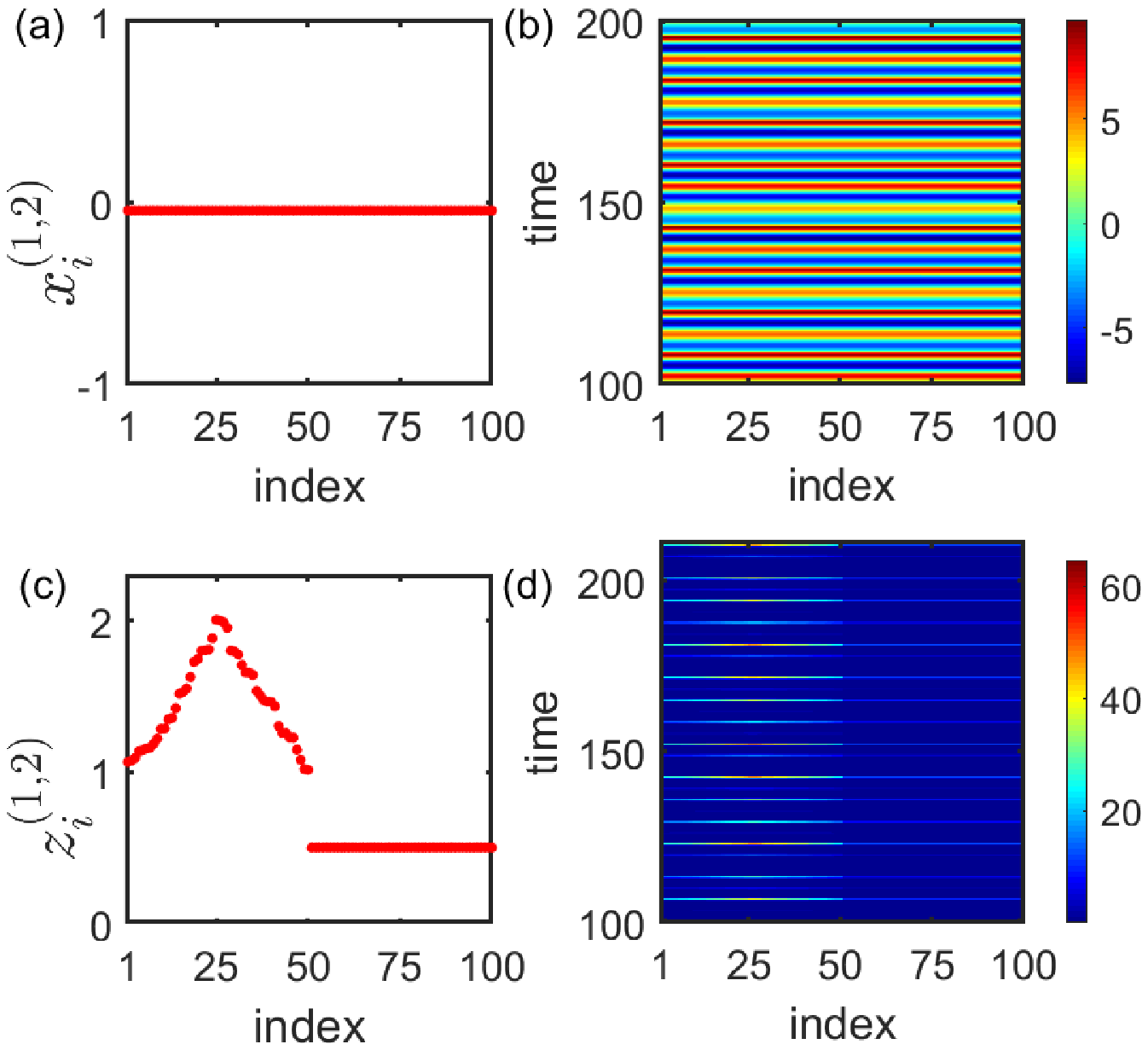}
		\includegraphics[width=3.5in,height=1.45in]{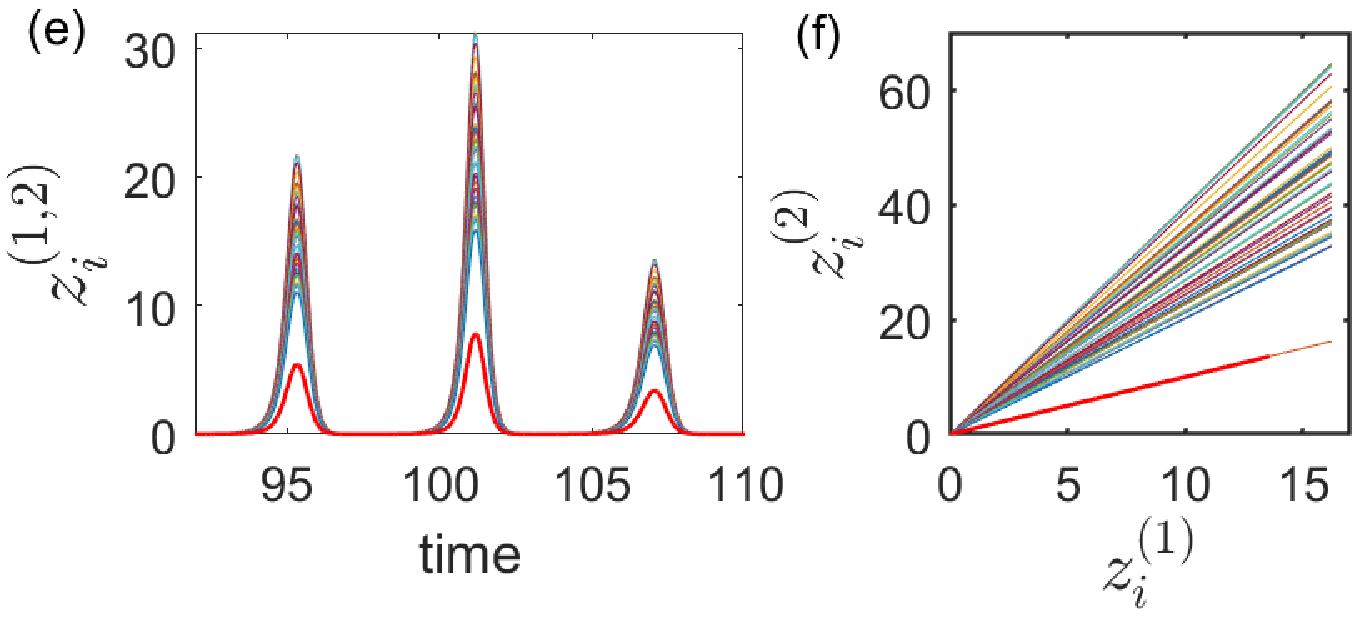}
		\caption{(Color online)  {\bf Multiplex network of R\"ossler oscillators.} Node indices 1-50 and 51-100, respectively, belong to Layer-1 and Layer-2. Self-coupling strengths $\varepsilon_1=1$ and $\varepsilon_2=1$, and the strength of cross-coupling link is $\kappa=-1$ \cite{saha2017coupling, saha2017cross}. For a choice of parameters $a=0.2$, $c=4.8$ and $b^{(2)}_i=b=0.2$, nodes in isolation exhibit chaotic dynamics. We perturb layer-1 by taking a distribution in $b^{(1)}_i\in(0.4,0.8)$, $i=1,\cdots,50$. The entire network perfectly synchronized even under the induced heterogeneity due to the presence of cross-coupling links from Layer-2 to Layer-1, when the $x$-variable shows complete coherence in amplitude and phase as shown in (a) and (b). The amplitude of $z$-variables in layer-1 are linearly scaled up by the amount of parameter mismatches, although the phase coherence is not disturbed as confirmed by a snap shot of $Z_i^{1,2}$ of all the nodes in two layers (c) and  a spatio-temporal plot (d).  (e) Time evolution of $Z_i^{1,2}$  shows amplified replica (multiple colors) of all the perturbed nodes against the identical nodes (red line), when only the amplitude is scaled up with complete phase coherence.
			(f) $Z_i^{1}$ vs $Z_i^{2}$ plot. Synchronization manifolds of the detuned nodes are rotated along the transverse direction against the CS manifold (red line) of the identical nodes. The rotation of the individual plot depends upon the ratio of mismatched parameters to the original (unperturbed) parameter. 
		}    \label{R_sigma_e1_rand}
	\end{figure}
	where $x^{(1,2)}_i, y^{(1,2)}_i$ and $z^{(1,2)}_i$ are the state variables; superscripts $1$ and $2$ denote Layer-1 and Layer-2, respectively.  $A^{(1,2)}_{ij}$ is the adjacency matrix of an individual layer. We use a rewiring strategy developed by Watts and Strogatz \cite{watts1998collective} to construct a random network for each layer independently, hence making two different network structures for the two layers where the inter-layer connections are one-to-one. The randomness of the two layers is defined by their degree distributions  ($\in$ [1,8]). The average degrees of Layer-1 and layer-2 are 6 and 5, respectively. 
	
	\par The intra-layer coupling for both the layers is defined by a diffusive self-coupling function $(y^{(1,2)}_j-y^{(1,2)}_i)$ with a coupling strength $\varepsilon_1$.
	One-to-one diffusive self-coupling links $(y^{(1,2)}_i-y^{(2,1)}_i)$ establishes the inter-layer connectivity between the adjacent nodes of the two layers with a strength $\varepsilon_2$. An isolated node exhibits chaotic dynamics by the choice of parameters as $a=0.2, c=4.8$ and  $b^{(1,2)}_i=b=0.2$ when all the nodes are identical. The identical multiplex network maintains 
	CS with local stability for coupling strength $\epsilon_1$ and $\epsilon_2$ decided by the MSF condition. Next, we introduce heterogeneity in all the nodes of Layer-1 with uniform distribution in  $b^{(1)}_i$ in a range $b^{(1)}_i\in (0.4,0.8)$ while all the nodes in Layer-2 are identical $b^{(2)}_i=b=0.2$, thereby destroying CS state in the multiplex network. Then, we apply our strategy of selective addition of directed cross-coupling links to each perturbed node of Layer-1 from the adjacent nodes of Layer-2.  The directed cross-couplings are linear and diffusive as defined by $(z^{(2)}_i-z^{(1)}_i)$  and added to the dynamics of the $x_i^{(1)}$-variable with a strength $\kappa$.
	The choice of the coupling links is made by our proposed systematic rules \cite{saha2017coupling}, and the strength of  $\kappa$ is defined by the LFS condition [see Supplementary material \cite{supple} for details]. Globally stable synchrony of the whole multiplex network is then reestablished. Figure \ref{R_sigma_e1_rand} presents numerical results.
	
	\par Figures for CS in the identical multiplex network are not presented here. The $x_i^{(1,2)}$-variable for the two layers in the perturbed network shows CS in amplitude and phase in a snapshot in Figure~\ref{R_sigma_e1_rand}(a) and a spatio-temporal plot in Figure~\ref{R_sigma_e1_rand}(b). It is also found true for  $y_i^{(1,2)}$ dynamics in all the nodes in two layers in the perturbed multiplex network (not shown here). However, the amplitudes $z_i$ of 50 nodes in Layer-1 are seen distributed in Figure~\ref{R_sigma_e1_rand}(c) while the other 50 nodes of layer-2 remain in CS. The spatio-temporal plot in Figure~\ref{R_sigma_e1_rand}(d) shows phase coherence among all the nodes. Looking at the temporal dynamics of the nodes in both the layers in Figure~\ref{R_sigma_e1_rand}(e), it is clear that the amplitudes of the perturbed 50 nodes in Layer-1 are linearly scaled up by  $\frac{r^{(1)}_i}{r}$, where $r_i^{(1)}$ follows a distribution, which is reflected in the amplitude distribution of the first $50$ nodes of Layer-1 in  Figure~\ref{R_sigma_e1_rand}(c). Due to induced distributed heterogeneity, the GS  manifold of the detuned nodes, $z_i^{(2)}$ vs. $z_i^{(2)}$, in Figure~\ref{R_sigma_e1_rand}(f), shows a variation in rotation angle of individual plots along the transverse direction of the CS manifold (red line). However, the phase coherence is not disturbed as usual by the distribution of parameter $b_i^{(1)}$ as revealed in Figures~\ref{R_sigma_e1_rand}(e)-(f).

	\section{Conclusion} \label{con}
	An essential question in multiplex networks is how to prevent loss of synchrony against parameter perturbation (or drifting)? We addressed the question with two examples of two-layered multiplex networks with a suggestion of adding selective linear diffusive cross-coupling links to the perturbed nodes in one layer from the adjacent nodes of the unperturbed layer without deleting any link from the original network. The selection of the coupling functions (self- and cross-coupling links) is most important for the multiplex network that follows a systematic rule formulated from the LFM of the dynamics of the nodes and an analytical procedure based on the LFS measure. This selection procedure is primarily elaborated in our earlier work \cite{saha2017coupling} and briefly explained in the Supplementary material. 
	\par An identical multiplex network with self-coupling maintains locally stable CS following the MSF conditions. Under parameter perturbation in any node of one layer, CS is broken when selective cross-coupling links restore synchrony as a globally stable GS state in the multilayer network. More categorically, all the nodes attain a globally stable CS except the perturbed nodes that emerge into a GS state. The amplitude of some of the state variables of the perturbed nodes shows a linear amplitude response to the amount of parameter perturbation in the GS state; the perturbed nodes remain phase coherent preventing a frequency drift, and it is independent of the size of the perturbation. This approach enables us to design resilient multiplex networks. 
	\par To demonstrate the efficacy of our proposal, we consider two exemplary multiplex networks with different topologies and different dynamical systems. As a first example, we use a multiplex network of two layers with the slow-fast FHN system representing the nodal dynamics and consider unidirectional chemical synapses that interact between the nodes of both layers, representing the nonlinear coupling between the nodes. We use the nonlocal coupling topology of the individual layers. CS is established with an assumption of identical nodes and deriving the critical coupling strength using the MSF formalism in the multiplex network. One of the parameters in one node in a layer is then perturbed, and we define the appropriate cross-coupling function as decided from the FHN dynamics and add to the perturbed node from the unperturbed layer. A globally stable GS state is realized when the stability conditions are derived using the LFS measure. 
	
	\par Most importantly, none of the original links of the network is deleted in contrast to what usually happens during the rewiring of links as used by others for enhancing synchrony in networks. In the unperturbed state, the cross-coupling links sleep without affecting the original dynamics and the CS state and awake or become active when a drift in parameters of the nodes appears. We successfully tested the results as true for induced heterogeneity in all the nodes in one layer when one cross-coupling link is to be added to each node from the immediate adjacent node of the other layer. Such a design of cross-coupling links can be adopted {\it a priori} for realizing a resilient multiplex network. We confirm our results with the second example of a multiplex network, where two layers of random networks with multiplexing are used, and a chaotic R\"ossler oscillator represents the nodal dynamics. Our proposed framework of additional cross-coupling links is thus independent of the dynamics of the nodes and the topology of the layers in a multiplex network.

	\ifCLASSOPTIONcompsoc
	\section*{Additional information}
	\else
	\section*{Additional information}
	\fi
	Selection of coupling profile from the linear flow of FitzHugh Nagumo and R\"ossler system are provided in the Supplementary material, Secs. 1.1 and 1.2, respectively. The stability conditions of CS of the FHN system (1) in the multiplex network are derived using the MSF in the supplementary material, Sec. 2.
	For Matlab codes, please visit, 
	https://github.com/ecesuman06/cross-coupling-multiplex.

	\ifCLASSOPTIONcompsoc
	\section*{Acknowledgments}
	\else
	\section*{Acknowledgment}
	\fi
	
	My immense gratitude to Dr. Syamal K. Dana for fruitful discussions, suggestions and help  preparing the article. I also acknowledge Dr. Ritwika Mondal for fruitful discussions and help in MSF derivation. I am very grateful to the anonymous reviewers for their critical comments and suggestion that helped me improve the quality of the work.

	\bibliographystyle{ieeetr}
	\bibliography{reef}{}

	\begin{IEEEbiography}[{
			\includegraphics[width=1in,height=1.25in,clip,keepaspectratio]{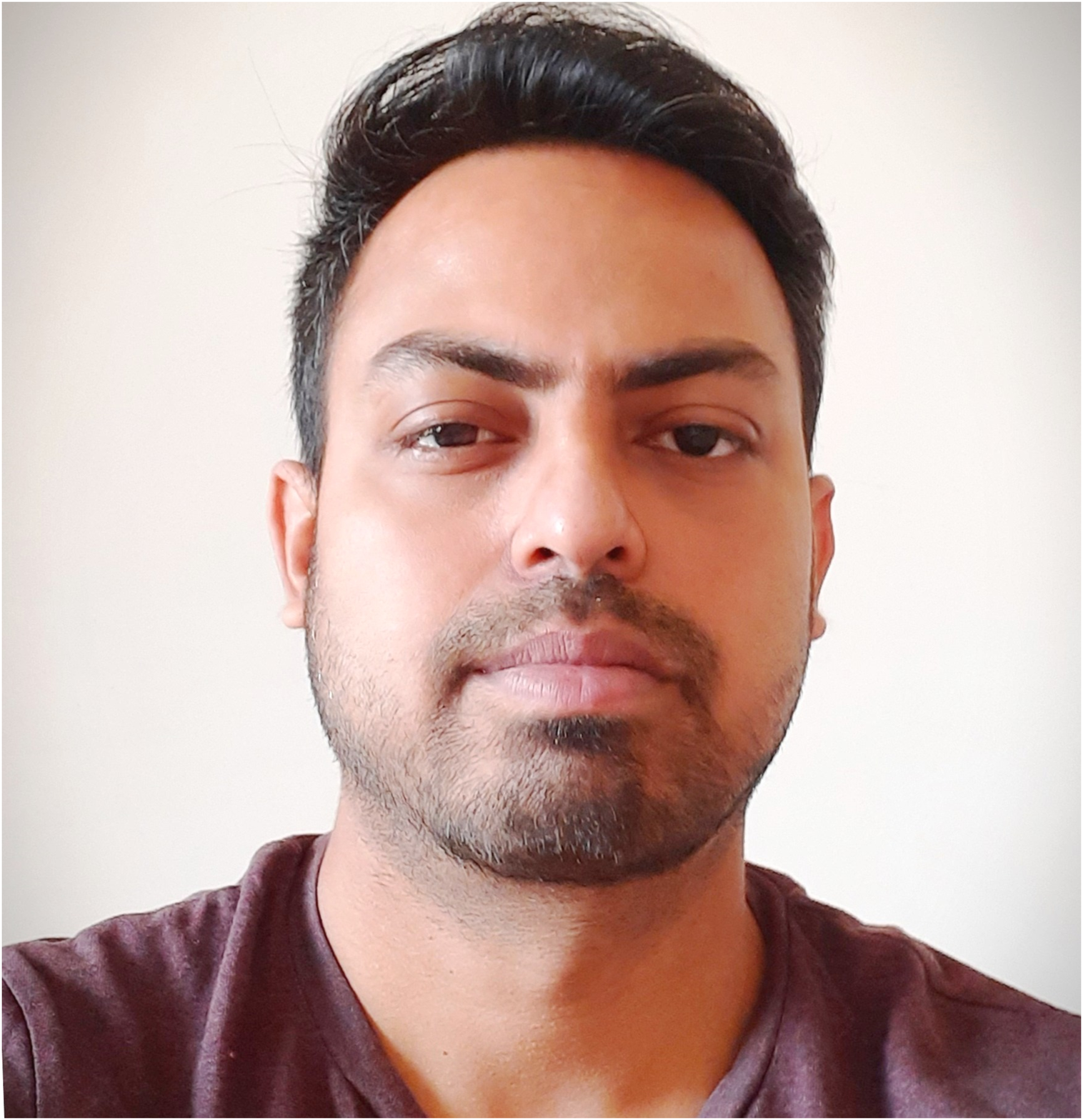}
		}]{Suman Saha}
		received the B.TECH., and M.TECH. degrees in Electronics and Communication Engineering both from West Bengal University of Technology, Kolkata, India, in 
		2006 and 2010, respectively. He completed his Ph.D. at Jadavpur University, Kolkata-700032, India, in 2020. His research interests include nonlinear dynamics, complex network, synchronization in dynamical systems and collective behaviors- cluster and chimera states, extreme events in neuronal systems, computational biology, the role of network structure and initial infectivity pattern in progression of infectious disease, predicting complex signal using machine learning, and neuroscience: cognition and brain dynamics.
	\end{IEEEbiography}

\end{document}


\title{Supplementary: Resilience in multiplex networks by addition of cross-repulsive links}

\author{Suman Saha,~\IEEEmembership{Project Scientist-I}
              
\IEEEcompsocitemizethanks{\IEEEcompsocthanksitem S. Saha is with National Brain Research Center, NH-8, Manesar, Gurugram-122051, India\protect\\
E-mail: ecesuman06@gmail.com
}
\thanks{}} 

\markboth{IEEE Transactions on Network Science and Engineering}%
{Saha\MakeLowercase{\textit{}}: Resilience in multiplex networks by addition of cross-repulsive links}

%

\maketitle

\IEEEdisplaynontitleabstractindextext

\IEEEpeerreviewmaketitle

\section{Selection of coupling profile from the linear flow of a system}\label{appendA}
The particular selection of a coupling profile (self-coupling and cross-coupling links) for the Lorenz system can really be made in a systematic manner from the linear flow matrix (LFM) of the system, and we can frame a set of general coupling conditions for the selection of a coupling profile for a dynamical system,
\begin{align}
	\dot{\bf x}=g({\bf x})= {\bf F x}+L({\bf x})+{\bf C}
\end{align}
$g:$$\mathbb{R}^n$$\rightarrow$$\mathbb{R}^n$ is the flow of the system, ${\bf F}$ is the LFM ($n \times n$) matrix, $n = 2,3$, respectively for our example systems, FHN and R\"ossler. $L({\bf x})$ represents the nonlinear functions,
and {\bf C} is a constant matrix. The coupling profile defines the choice of self-coupling and cross-coupling matrices, $H_s$ and $H_c$, respectively, from the LFM as elaborated by the following examples.

\subsection{FitzHugh Nagumo (FHN) model}
\noindent Dynamical equation of FHN \cite{fitzhugh1961impulses} systems
\begin{align} \label{eq1}
	&\dot{x} = \; x -x^3-{y} +I\nonumber\\ 
	& \dot{y} = \; rx-by 
\end{align}

\noindent For the FHN system \eqref{eq1},

\[ {\bf F}=
\begin{bmatrix}
	1      &  -1 \\
r      & -b
\end{bmatrix}, \quad
L({\bf x})=
\begin{bmatrix}
	-x^3       \\
	0     
\end{bmatrix},\quad
{\bf C}=
\begin{bmatrix}
	I       \\
	0     
\end{bmatrix}
\]

\noindent By an inspection of LFM ({\bf F}) of the FHN system, we a suggest that
 self-coupling is necessary involving the $x$ variables of the coupled systems since the element $F_{11}=1$ is positive. While a nonzero element exists in the off-diagonal element $F_{12}$ ($=-1$) that is connected to the $\dot{x}$ dynamics. Hence a cross-coupling function involving the $y$ variables are added the evolution equation of $x$. The coupling profile is defined by the self- and cross-coupling matrices $H_s$ and $H_c$, respectively,  
\[
H_s=
\begin{bmatrix}
	1      &  0 \\
	0     & 0
\end{bmatrix}, \quad
H_c=\begin{bmatrix}
	0      &  -1 \\
	0      &  0
\end{bmatrix}. \quad
\]

Note here that this particular manner of selection the coupling functions from an information of the LFM is not arbitrary, but based on a general principle prescribed in details in \cite{saha2017coupling}.
\noindent For our example of two coupled FHN systems, the LFM of the system thus suggests one self-coupling involving $x$ variables of the systems and to be added to evolution equation of $x$ and it is most appropriate for realizing complete synchronization (CS) \cite{saha2017coupling} and one directed
cross-coupling link defined by a coupling function involving
the $y_{1}$ variable and to be added to the $x_1$ evolution equation as shown in Eq. \eqref{eq2}, since
the nonzero element is connected to the $y_{1}$ variable in $\dot{x_1}$. This additional cross-coupling is a necessary condition as decided by the Lyapunov function stability (LFS), for attaining global stability of CS. Hence we write the coupled system,
\begin{align}	\label{eq2}
	&\dot{x}_{1} = \; x_1 -x_{1}^3-{y}_{1} +I+ \varepsilon(x_{2}-x_{1})+\kappa(y_{2}-y_{1})\nonumber\\ 
	& \dot{y}_{1} = \; r_{1}x_{1}-by_{1}  \nonumber\\
	&\dot{x}_{2} =  \; x_2 -x_{2}^3-{y}_{2} +I+ \varepsilon(x_{1}-x_{2}) \nonumber\\
	& \dot{y}_{2} = \; r_{2}x_{2}-by_{2} 
\end{align}
where $(x_{1,2},y_{1,2})$ are state variables,  $\varepsilon$ is self-coupling strength and $\kappa$ is the cross-coupling strength. 
\noindent The conventional diffusive coupling function $(x_{2,1}-x_{1,2})$ is stated as self-coupling when it is added to the evolution equation of the same variables $x_{1,2}$. Mismatch is introduced in parameter by taking $r_1$ and $r_2$ different. Furthermore, a directed cross coupling link $(y_2-y_1)$ is added to the dynamics of $x_1$ (e.g. from node-2 of layer-2 to node-2 of layer-1 in the multiplex network in the main text).

\noindent Now, we perform the Lyapunov function stability (LFS) analysis to find the value of cross-coupling strength ($\kappa$), between the two systems (e.g. node-2 of layer-1 and node-2 of layer-2 in the multiplex network). We assume here that each layer attains CS before adding the cross-coupling for an appropriate choice of $\epsilon$ and hence reduced to one system.
%
%
\noindent The error functions $\bf e$=$[e_x,e_y]^T=[x_1-x_2,y_1-y_2]^T$ of the systems Eq.~\eqref{eq2} evolves, as
\begin{align}
\dot{e}_x = &\;  e_x-\frac{e_x^3}{4}-\frac{3}{4}e_xe_p^2-e_y-2\varepsilon e_x-\kappa e_y\nonumber\\
\dot{e}_y = &\; re_x -be_y \label{s5}
\end{align}
\noindent where, $e_p=x_1+x_2$ so that $x_1^3-x_2^3=\frac{e_x}{4}(e_x^2+3e_p^2)$.
For a global stability of the synchronous state $(e_x=0,e_y=0)$, we consider a Lyapunov function, $V(e)=\frac{1}{2}e_x^2+\frac{1}{2}e_y^2$.
We first check the stability of $x_1=x_2$, separately, by defining a Lyapunov function, $V'(e_x)=\frac{1}{2}e_x^2$ when its time derivative is
\begin{align}
\dot{V'}(e_x)=&-e_x^2 \left( \frac{3}{4}e_p^2-1+2\varepsilon_2 \right)-\frac{e_x^4}{4} -(1+\kappa)e_ye_x.\label{s6}
\end{align}
\noindent $\dot V'$($e_x)<0$, when  $\kappa=-1$ and $\frac{3}{4}e_p^2-1+2\varepsilon\geq 0$. So, the cross-coupling link with $\kappa=-1$ will remove the contribution of  $e_y$, which establishes the global stability between the two nodes.
%
\noindent This implies, synchronization manifold for $x_1=x_2$ and $y_1=y_2$ state is asymptotically stable as $t\rightarrow\infty$. Assuming identical  systems ($r_1=r_2$), the error dynamics  is found to be $\dot{e}_y=-be_y$, when
\begin{equation} \label{s9}
\dot{V}(e_x,e_y)=-\frac{e_x^4}{4} - b e_y^2 <0, \quad \varepsilon\leq \frac{1}{2} \quad \text{and} \quad \kappa=-1.
\end{equation}
%
%
%
\noindent Accordingly, the Lyapunov function is redefined in terms of the modified error functions in case of non-identical systems ($r_1\neq r_2$), whose time derivative can be rewritten as,
\begin{equation}
\dot{V}^{*}(e_x,e_y^*)=-\frac{e_x^4}{4}-be_y^{*2}<0 \label{s13}
\end{equation}  
\noindent where, 	$e_y^*=y_1\frac{r_1}{r_2}-y_2$, which ensures globally stable synchronization manifold in the mismatched systems.

\subsection{R\"ossler system}
The dynamical equation of R\"ossler system is
\begin{align} \label{ros}
	&\dot{x}=-y-z \nonumber \\
	&\dot{y}= x + ay \\
	&\dot{z}= b+xz-cz \nonumber 	
\end{align}

\noindent The system \eqref{ros} can be separated into its LFM, nonlinear function $L(x)$ and constant ${\bf C}$ matrices,
\[ {\bf F}=
\begin{bmatrix}
	0 &  -1 &  -1 \\
	1 &   a &   0 \\
	0 &   0 &  -c
\end{bmatrix}, \quad
L({\bf x})=
\begin{bmatrix}
	0       \\
	0     \\
	xz
\end{bmatrix},\quad
{\bf C}=
\begin{bmatrix}
	0       \\
	0 \\
	b    
\end{bmatrix}.
\]

\noindent The self- and cross-coupling matrices, respectively, can be defined \cite{saha2017coupling} from the elements  of the LFM,
\[
H_s=
\begin{bmatrix}
	0     &  0 & 0 \\
	0     & 1  & 0 \\
	0     & 0  & 0 
\end{bmatrix}, \quad
H_c=\begin{bmatrix}
0     &  0 & -1 \\
0     & 0 & 0 \\
0     & 0 & 0 
\end{bmatrix}. \quad
\]
 A self-coupling function involving $y_{1,2}$ variables is necessarily to be added to the evolution equations of $y_{1,2}$ to establish locally stable CS since the diagonal element in the second row $F_{22}=a$ is a constant. An alternative choice of a self-coupling coupling function $x_{1,2}$ can be made since $F_{11}=0$ that is to be added to evolution equations of $x_{1,2}$, however, it cannot realize CS for larger coupling strength beyond the critical value. For an intermediate range of large coupling, it is known that CS in the coupled R\'ossler systems breaks down while the LFM proposed self-coupling function involving $y_{1,2}$ variables has no such limit beyond the critical coupling. And the self-coupling involving $z_{1,2}$ is feasible to realize CS as also known in the literature, however, confirmed here by the diagonal element value $F_{33}=0$. 
\noindent Therefore, for two coupled
R\"ossler system, the coupling profile is defined by as a bidirectional
self-coupling that involves $y_{1,2}$ variables and a cross-coupling that
involves $z_{1,2}$ variables and added the dynamics of $\dot{x}_1$. The
dynamical equations of two-coupled R\"ossler systems,
\begin{align} \label{rosc}
	&\dot{x}_1=-y_1-z_1 + \kappa (z_2-z_1)\nonumber \\
	&\dot{x}_2=-y_2-z_2 \nonumber \\
	&\dot{y}_{1,2}= x_{1,2} + ay_{1,2}+\varepsilon(y_{1,2}-y_{2,1}) \\
	&\dot{z}_{1,2}= b_{1,2}+x_{1,2}z_{1,2}-cz_{1,2} \nonumber 	
\end{align}
The error dynamics  is
\begin{align} \label{rose}
	&\dot{e}_x=-e_y-e_z + \kappa (z_2-z_1)\nonumber \\
	&\dot{e}_y= e_x + (a-2\varepsilon)e_y \\
	&\dot{e}_z= (b_1-b_2) + x_1z_{1}-x_2z_2-ce_z \nonumber 	
\end{align}
We determine the stability of $x_1 = x_2$ and $y_1 = y_2$  first. This
involves the construction of a Lyapunov function $V(e_x,e_y)$,
which is positive definite function, $V(e_x,e_y) = \frac{1}{2}e_x^2+\frac{1}{2}e_y^2$.
The time derivative of the Lyapunov function,  $\dot{V}(e_x,e_y) = -(1 + \kappa)e_xe_z + (a - 2\varepsilon)e_y^2$. 

For $\varepsilon >a/2$ and $\kappa = -1$,  $\dot{V}(ex,ey)<0$ is negative semidefinite, since we get $\dot{V}(e_x,e_y ) = 0$ if $e_y = 0$ and for any values
of $e_x$. However, by using the LaSalle invariance principle \cite{la1966invariance},
the set $S = \{e_x,e_y\}$ does not contain any trajectory except
the trivial trajectory $(e_x,e_y) = 0$, as $\dot{e}_y \neq 0$ if $e_x \neq 0$. As
a result, the trajectory will not stay in the set $S$. So the
synchronization manifold of $x_1 = x_2$ and $y_1 = y_2$ is
asymptotically stable with $t\rightarrow \infty$ above a critical value of $\varepsilon_2$ and $\kappa=-1$.

Now, in case of nonidentical oscillators ($b_1 \neq b_2$), the stability in $e_x$ and $e_y$ is undisturbed by the induced
heterogeneity when $ \dot{V}(e_x,e_y) \leq 0$ remains valid. Next, we
check the error dynamics $e_z$, which will be revised as,
$\dot{e_z^*}=-ce_z^*(1-x_1/c)$, where, $e_z^*=z_1\frac{b_1}{b_2}-z_2$. Hence, using the LaSalle invariance principle \cite{la1966invariance} the coupled R\"ossler system Eq. \eqref{rosc} is globally
synchronized, provided $c >| x_1 |$. In fact, it shows an error to
the limit of $10^{-4}$.

\section{MSF of the Multiplex network: FHN system} \label{msfmn}

	\begin{align} \label{eqI50}
	\dot{x}_i^{l} = & F_x(x^{l}_i,y^{l}_i) + \frac{\varepsilon_1 (x^{l}_i-V_s)}{2P}\sum\limits_{j=i-P}^{i+P}  H(x^{l}_j) \nonumber\\
	&+\varepsilon_2 Q(x_i^{(n,l)}-x_i^{(l,n)})+\kappa ~ C_{i,i}^{(n,l)}(y_i^{n}-y_i^{l}), \nonumber \\
	\dot{y}_{i}^{l} = &\;  F_y(x^{l}_i,y^{l}_i,r^{l}_{i}), 
\end{align}

The stability conditions of CS state of the system \eqref{eqI50} are derived using the MSF \cite{pecora1998master} approach to determine the critical values of  $\varepsilon_1$ and $\varepsilon_2$, in absence of cross-coupling links ($\kappa=0$). 
For identical case, $r^{(l)}_1$=$r^{(l)}_2$=$\cdots$=$r$=$1$, a stable CS state is easily realized with no necessity of cross-coupling links ($\kappa=0$). For analytical convenience and simplicity, we denote,

{\footnotesize
	\begin{align*}
		\hspace{-5pt} &z^{(l)}_i=[x^{(l)}_i,y^{(l)}_i]^T;
		f(z^{(l)}_i)=[F_x(x^{(l)}_i,y^{(l)}_i),F_y(x^{(l)}_i,y^{(l)}_i,r)]^T;\\
		&z^{(l)}=[z^{(l)}_1, z^{(l)}_2,..., z^{(l)}_N]^T;
		\bar{f}(z^{(l)})=\left[f(z^{(l)}_1), f(z^{(l)}_2),...,f(z^{(l)}_N)\right]^T; \\
		&\mathbf{z}=[z^{(1)},z^{(2)}]^T;~~
		F(\mathbf{z})= \left[\bar{f}(z^{(1)}),\bar{f}(z^{(2)})\right]^T.
	\end{align*}
}
Eq.~\eqref{eqI50} is then expressed as,
\begin{align} \label{msf1}
	\dot{\mathbf{z}}&= F(\mathbf{z})-\varepsilon_1(\mathcal{L}^{L} \otimes H)\mathbf{z}-\varepsilon_2(\mathcal{L}^{l} \otimes Q)\mathbf{z},
\end{align}
where, $\mathcal{L}^{L}$ and $\mathcal{L}^{l}$ denote the supra-Laplacian of intra- and inter-layer connectivity matrices. A detail description can be found in \cite{tang2019master,gomez2013diffusion,sole2013spectral}. 
To determine the MSF for Eq.~\eqref{eqI50}, we linearize system \eqref{msf1} at $\mathbf{1}_{2} \otimes \mathbf{1}_N \otimes {s}$, where $s$ represents the CS state, satisfying 
$\dot{s}=f(s);$
$\mathbf{1}_{2}$ stands for 2-dimensional vector with all entries being 1. Thereby, we obtain the variational equation,
\begin{eqnarray*} \label{msf2}
	\dot{\xi}=[ I_{2 \times N}  \otimes Df(s) - \varepsilon_1 (\mathcal{L}^{L} \otimes \bar{H}) -\varepsilon_2 (\mathcal{L}^{l} \otimes Q)]\xi
\end{eqnarray*} 
where, $\xi=\mathbf{z}-\mathbf{1}_2 \otimes \mathbf{1}_N \otimes s ~\mbox{and}~ I_{2 \times N}$ is the identity matrix of order $2 \times N$. After a mathematical manipulation, we obtain $\bar{H}(s)$ at synchronous state $s$,
\begin{equation}
	\bar{H}(s)=
	\begin{pmatrix}
		\varepsilon_1 H(s) & 0 \\
		0 & 0
	\end{pmatrix} +
	\begin{pmatrix}
		\zeta^{(l)} H'(s) & 0 \\
		0 & 0
	\end{pmatrix},
\end{equation}  
$H'(s)$ is first order derivative of $H(s)$.  $\mathcal{L}^{L}$ and $\mathcal{L}^{l}$ are assumed symmetric and commutative.
After diagonalization and  decoupling, we have obtained the  MSF associated to system \eqref{eqI50} 
\begin{align}
	\dot{\omega} = & \; [  Df(s) - \alpha \bar{H} -\beta Q] \omega.\label{msf}
\end{align} 
where $\alpha=\varepsilon_1 \lambda $, $\beta=\varepsilon_2 \mu$  and $\lambda$, $\mu$ are real eigenvalues of $\mathcal{L}^{L}$ and $\mathcal{L}^{l}$, respectively, where $\lambda^2+\mu^2\neq 0$.

The Lyapunov exponents are determined by the following linearized equation with respect to the reference trajectory for the synchronous state $s(t)$ with initial condition $U(0)$:
$\dot{U}=J(s(t))U$, where $J$ is the Jacobian matrix of $f(s)$. Let $v_i(0)(i=1,2,\cdots,4N)$ is the orthogonal vector of $U(0)$. The Lyapunov exponents are defined as, 
$$\varrho_i=\lim_{t\to\infty}\frac{1}{t}\text{ln}||U(t){v}_i(0)||.$$
The multiplex network shows locally stable CS when $LLE(\alpha,\beta)<0$, where $LLE=max(\varrho_i)$.

\section{Matlab codes}
For matlab codes please visit 
https://github.com/ecesuman06/cross-coupling-multiplex

\bibliographystyle{ieeetr}
\bibliography{reef}{}